\documentclass[twocolumn,aps,prb,superscriptaddress,amssymb,showpacs]{revtex4-1}
\usepackage[dvips,final]{graphicx}
\begin{document}

\title{Enhancement of the ferromagnetic order of graphite after sulphuric acid treatment}

\author{J. Barzola-Quiquia}
\author {W. B\"ohlmann}
\author{P. Esquinazi}\email{esquin@physik.uni-leipzig.de}
\author{A. Schadewitz}
\author{A. Ballestar}
\author{S. Dusari}
\affiliation{Division of Superconductivity and Magnetism, Institut
f\"ur Experimentelle Physik II, Universit\"{a}t Leipzig,
Linn\'{e}stra{\ss}e 5, D-04103 Leipzig, Germany}
\author{L. Schultze-Nobre}\altaffiliation[Present address: ]{Helmholtz Centre for Environmental
Research-UFZ, Permoserstr 15, D-04318 Leipzig, Germany}
\author{B. Kersting}
\affiliation{Institut f\"ur Anorganische Chemie, Universit\"{a}t
Leipzig, Johannisallee 29, D-04103 Leipzig, Germany}


\begin{abstract}
We have studied the changes in the ferromagnetic behavior of
graphite powder and graphite flakes after treatment with diluted
sulphuric acid. We show that this kind of acid treatment enhances
substantially the ferromagnetic magnetization of virgin graphite
micrometer size powder as well as in graphite flakes. The
anisotropic magnetoresistance (AMR) amplitude at 300~K measured in
a micrometer size thin graphite flake after acid treatment reaches
values comparable to polycrystalline cobalt.
\end{abstract}


\maketitle

The possibility to trigger magnetic order in metal-free graphite
by lattice defects or non-magnetic adatoms like hydrogen attracts
the interest of  basic and applied research fields \cite{yaz10}.
Recently published element specific x-ray magnetic circular
dichroism (XMCD) measurements at the carbon K-edge in graphite
samples \cite{ohldagl,ohldagnjp} clearly demonstrated that the
reported magnetic order in metal-free graphite
\cite{pabloprb02,pabloprl} is not related to magnetic impurities.
Published results from different groups indicate that vacancies
\cite{xia08,yan09,ram10} and/or hydrogen
\cite{duplock04,ohldagnjp} can trigger this phenomenon, making the
graphite structure the archetype of defect induced magnetism
(DIM), a phenomenon that is being now found in nominally
non-magnetic oxides \cite{pot08,kha09} as well as in Si-based
samples \cite{liu11} for example. In case of graphite magnetic
order has been also achieved by a pulsed arc ignited between two
graphite electrodes in ethanol \cite{par08}. In general, however,
the  obtained yield remains small, partially because of the
necessary delicate balance between a defect density of the order
$\sim 5\%$, their positions and the lattice structure. For
applications it is necessary therefore to find simpler methods to
transform graphite  in a magnet with reasonable high
magnetization.

XMCD results in untreated as
 well as in proton irradiated bulk pyrolytic graphite provide clear hints for
the influence of hydrogen atoms, especially a spin polarization
splitting near the Fermi level \cite{ohldagnjp} in qualitative
agreement with theoretical work published previously
\cite{duplock04}. The aim of the here reported study was to
 find a simpler method to trigger magnetic order in  graphite samples of mesoscopic size by
 including hydrogen at least in the near surface region
 without destroying its lattice structure or including further defects or contaminants.
One possibility is to treat carbon with sulfuric acid leading to a
hydrogen doping in the graphite structure.

\begin{figure}[]
\begin{center}
\includegraphics[width=1\columnwidth]{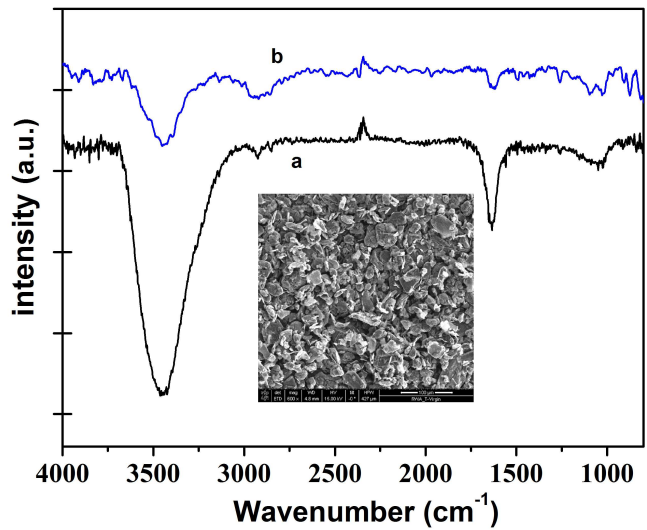}
\caption[]{Infrared absorption spectra  of the graphite powder
before (a) and after (b) acid treatment (sample S3). The inset
shows a scanning electron microscope picture of the used powder.
The scale bar indicates $100~\mu$m length.}\label{irs}
\end{center}
\end{figure}
\begin{figure}[]
\begin{center}
\includegraphics[width=0.8\columnwidth]{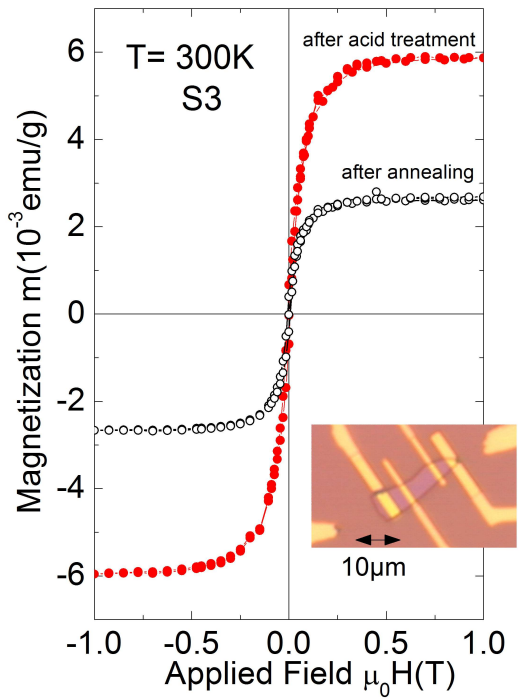}
\caption[]{Magnetization vs. applied magnetic field for sample S3
after acid treatment  and after annealing two hours in vacuum at
250$^\circ$C. The inset shows an optical microscope picture of the
measured graphite flake with the contact Au
electrodes.}\label{ann}
\end{center}
\end{figure}

Two kinds of graphite samples have been used in this work, namely
an ultrapure graphite powder (impurity content $< 1~$ppm)
consisting of platelet-like grains of average size $10 \times 10
\times 3\mu$m$^3$ (see inset in Fig.~\ref{irs}) and a micrometer
size graphite flake of thickness $\simeq 45~$nm (see inset in
Fig.~\ref{ann}). The influence of the sulphuric acid on the powder
has been measured using a superconducting quantum interferometer
device (SQUID) and by transport measurements in the case of the
graphite flake, especially the anisotropic magnetoresistance (AMR)
at different angles between field and current for fields applied
parallel to the graphene planes of the sample.

\begin{table}[ht]
\centering
\begin{tabular}{|ccc|} \hline Sample & H$_2$O & H$_2$SO$_4$ \\   S0 & 10~ml & 10~ml \\
 S1 & - & 20~ml \\  S2& 20~ml & - \\ S3 & 5~ml
& 15~ml \\ S4 & - & - \\ \hline\end{tabular} \caption{Powder
samples prepared with different concentrations of sulphuric
acid.}\label{mytable}
\end{table}

Each sample for the magnetization measurements was prepared mixing
100~mg of graphite powder into 20~ml of a liquid with sulphuric
acid and water with a given concentration, see
Table~\ref{mytable}. The suspension was continuously stirred at
room temperature for 24~h. After stirring the obtained material
was recovered by filtration and washed with distilled water to
remove residual traces of acid and dried at 100~$^\circ$C
overnight for its characterization with the SQUID and infrared
spectroscopy (IRS). For the SQUID measurements we used 10~mg of
the prepared unpressed powder. For the IRS measurements we used
the KBr method, i.e. 1~mg of the previously prepared graphite
powder and 100~mg of KBr were mixed and pressed to form a tablet.
The IR spectra were performed in transmission on a Perkin Elmer
spectrometer. The micrometer size sample for the AMR measurements
was obtained from a high oriented pyrolitic grapite (HOPG) sample
by the rubbing method (for details see Ref.~\onlinecite{bar08})
and ohmic contacts were prepared using electron beam litographie
method.


Figure~\ref{irs} shows the  infrared absorption spectra of the
graphite powder as received and after acid treatment. The
comparison of the two spectra clearly shows the influence of the
acid treatment. The increase in the intensity of the peak at
2943~cm$^{-1}$ can be assigned to formed -C-H- stretching bands.
Furthermore, in the region at 1650~cm$^{-1}$ a medium band is
observed, which generally belongs to C=C double bonds in aromatic
rings \cite{sil81}. The intensity of this stretching band
decreases after the treatment with sulphuric acid demonstrating
that partial hydrogen is embedded into the near surface region of
the graphite structure. Similar results were obtained in
Ref.~\onlinecite{obr09} in activating processes of carbon black
and carbon nanotubes with a mixture of concentrated H$_2$SO$_4$ +
HNO$_3$ in an ultrasonic bath. The activation leads to the
intensification of the band due to -OH vibration and to the
appearance of many other bands such are those around
2920~cm$^{-1}$ and between 1723 and 1090~cm$^{-1}$ originating
from vibration of oxygen functionalities. we note however, that
according to the last XMCD results oxygen should not play any
prominent role in triggering the magnetic order in graphite
\cite{ohldagnjp}.

\begin{figure}[]
\begin{center}
\includegraphics[width=0.9\columnwidth]{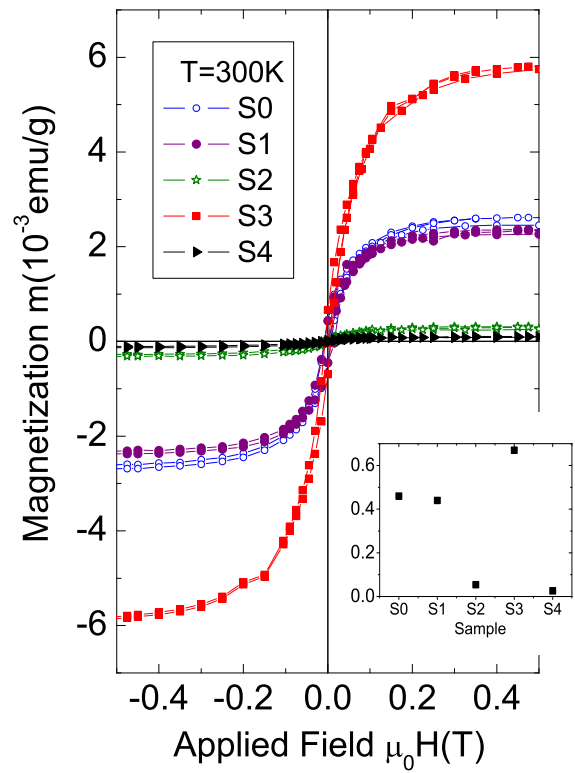}
\caption[]{Magnetization hysteresis loops for the differently
prepared samples, see Table~\protect\ref{mytable}. The inset shows
the remanent magnetization in the same units for the different
samples. The error in this measurement is less than the symbol
size.}\label{squid}
\end{center}
\end{figure}

Sample~S2 was treated only with water using exactly the same
procedure to check that it did not introduce any significant
amount of impurities. The magnetic moment of the sulphuric acid
alone and at room temperature did not show any ferromagnetic
signals (not shown). Figure~\ref{squid} shows the magnetization
hysteresis loops at room temperature for the differently prepared
powder samples after subtraction of the linear diamagnetic
background. The magnetization was estimated taken the whole mass
of the powder sample, which was in all cases 10~mg. The results
presented in Fig.~\ref{squid} reveal a clear influence of the acid
on the ferromagnetic properties of the graphite powder. The small
values of the obtained magnetization indicate that only the near
surface region of the grains is affected by the acid treatment.
From the XMCD results \cite{ohldagnjp} we may assume an intrinsic
ferromagnetic magnetization at saturation $\gtrsim 15~$emu/g. With
the  obtained saturation magnetization values for sample S3 and
taking the average grain size we estimate an average magnetic near
surface region of thickness $\lesssim 2~$nm.

Figure~\ref{ann} shows the magnetization vs. applied field for
sample~S3 before and after annealing it two hours in vacuum at
250$^\circ$C.  We observe a clear decrease of the ferromagnetic
signal indicating that defects, in this case very probably
hydrogen is related to the origin of the ferromagnetic
enhancement.

An appropriate  method to check the existence of ferromagnetism in
a single micrometer small graphite sample is to measure the
transport properties. In this work we concentrate ourselves to
study the AMR, defined as the dependence of the magnetoresistance
on the angle between the direction of the electric current $I$ and
the magnetic field $H$, in this case applied always parallel to
the main area of the sample to keep any
 form-anisotropy or Lorentz-force contribution to the
magnetoresistance invariant.

\begin{figure}[]
\begin{center}
\includegraphics[width=1\columnwidth]{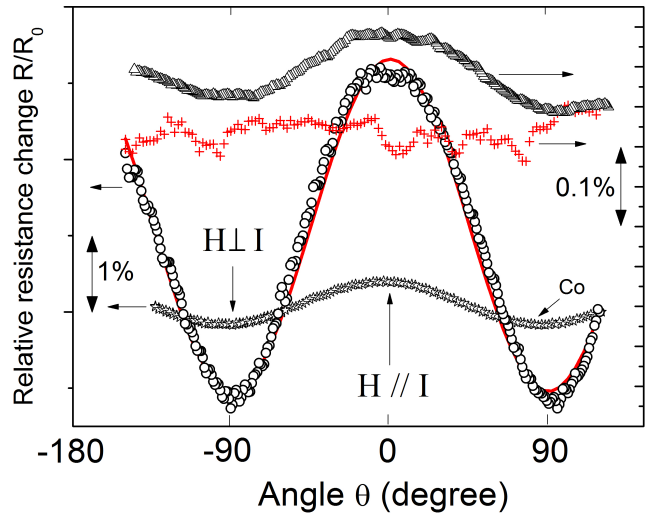}
\caption[]{Relative change of the electrical resistance vs. the
angle between the magnetic field ($\mu_0 H = 0.5$~T) and the
applied current for the graphite flake  at 300~K  before  ($+$,
right $y-$axis) and after acid treatment ($\bigtriangleup$, right
$y-$axis). ($\bigcirc$): data obtained after acid treatment at
30~K under same conditions (left $y$-axis). The continuous line
follows the $\cos^2(\theta)$ function, usual for the AMR of
polycrystalline magnetic samples. For comparison the AMR measured
in a Co polycrystalline film at 300~K is also shown ($\star$, left
$y-$axis). }\label{amr}
\end{center}
\end{figure}

Figure~\ref{amr} shows the magnetoresistance measured at a fixed
field of 0.5~T as a function of the angle between the input
current and the magnetic field. The graphite flake, see inset in
Fig.~\ref{ann}, in its virgin state and at 300~K shows negligible
AMR, whereas a clear AMR of $\simeq 0.1\%$ amplitude (defined
between the maximum and minimum) is measured after diluted acid
treatment. The sample was treated with an acid droplet, similar to
that used for sample S3. This amplitude can be compared to $\simeq
0.6\%$ obtained for a pure Co polycrystalline film of size $0.6
\times 0. 38~$mm$^2$ and 12~nm thickness, see Fig.~\ref{amr}.
Assuming that $\sim 2~$nm from the treated graphite flake becomes
ferromagnetic after the acid treatment and that the bulk of the
sample does not contribute to the AMR, a simple parallel resistor
model indicates that the real AMR amplitude of the ferromagnetic
near surface region should be at least 20 times larger, i.e. in
the 2\% range. The amplitude of the AMR increases decreasing
temperature as the data obtained at 30~K indicate, see
Fig.~\ref{amr}.
In ferromagnetic materials with elements with $d$-bands the AMR
effect is attributed to the larger probability for a $s-d$
scattering of electrons moving parallel to the applied field. Its
origin is related to: (1) the spin asymmetry of the $d-$band and
(2) to a finite $L-S$ coupling, which allows electrons to have a
spin-flip, enhancing in this way the scattering probability and
therefore the resistance. As pointed out in
Ref.~\onlinecite{jems08} the fact that magnetic graphite shows the
AMR phenomenon indicates both, a non-negligible $L-S$ coupling and
a spin splitting of the $p-$band. Although for carbon-based
systems a negligible $L-S$ coupling is expected due to the low
atomic number of carbon, our results indicate that this is not
true when a magnetic moment is originated at a vacancy or hydrogen
bonding. The results shown in Fig.~\ref{amr} indicate in fact a
giant AMR for hydrogen-mediated magnetic graphite, an effect that
supports the XMCD results of Ref.~\onlinecite{ohldagnjp}.

One of the authors (P.E.) acknowledges discussions with Prof. Dr.
S. Penad{\'e}s (CIC biomaGUNE, San Sebastian). This work is
supported by the Deutsche Forschungsgemeinschaft under contract
DFG ES 86/16-1. A.B. and S.D. are supported by the Graduate School
of Natural Sciences ``BuildMoNa" of the University of Leipzig.



%

\end{document}